\documentclass{article}
\usepackage[utf8]{inputenc}
\usepackage{times}
\usepackage{amsfonts}
\usepackage{amsmath,amssymb}
\usepackage{indentfirst}
\usepackage{wrapfig}
\usepackage{graphicx}
\usepackage{graphics}
\usepackage{caption}
\usepackage{subcaption}
\usepackage[toc,page]{appendix}
\usepackage{color}
\usepackage{cite}
\usepackage[inactive]{srcltx}
\usepackage[T1]{fontenc}
\usepackage{float}
\usepackage{hyperref}

\begin{document}
		\date{}
		\begin{center}
			{\Large\bf Performance analysis of continuous-variable quantum key distribution using non-Gaussian states}
		\end{center}
		\begin{center}
			{\normalsize L.S. Aguiar, L.F.M. Borelli, J.A. Roversi and A. Vidiella-Barranco \footnote{vidiella@ifi.unicamp.br}}
		\end{center}
		\begin{center}
			{\normalsize{Gleb Wataghin Institute of Physics - University of Campinas}}\\
			{\normalsize{ 13083-859   Campinas,  SP,  Brazil}}\\
		\end{center}
\begin{abstract}
In this study, we analyze the efficiency of a protocol with discrete modulation of continuous variable
non-Gaussian states, the coherent states having one photon added and then one photon subtracted (PASCS). 
We calculate the secure key generation rate against collective attacks using the fact that Eve's information
can be bounded based on the protocol with Gaussian modulation, which in turn is unconditionally secure. 
Our results for a four-state protocol show that the PASCS always outperforms the equivalent coherent states protocol 
under the same environmental conditions. Interestingly, we find that for the protocol using discrete-modulated PASCS,
the noisier the line, the better will be its performance compared to the protocol using coherent states.
Thus, our proposal proves to be advantageous for performing quantum key distribution in non-ideal situations.
\end{abstract}

\section{Introduction} 

%\label{Intro}
There has been increasing interest in continuous variables (CV) quantum key distribution (QKD) as an alternative to discrete 
variables (DV) QKD. An important step towards CV-QKD was the elaboration of a protocol using coherent states with added noise, 
that is, with Gaussian modulation of coherent states \cite{Grosshans02}. However, as original proposed, i.e. using direct reconciliation, 
the aforementioned protocol does not allow to perform QKD in transmission lines having losses greater than 50 $\%$ (3 dB). This limitation was 
soon overcome using two different approaches, either via post-selection \cite{Silberhorn02} or by reverse reconciliation procedures 
\cite{Grosshans02R}. The protocol using Gaussian-modulated coherent states and reverse reconciliation was experimentally implemented 
with pulses containing a few hundred photons \cite{Grosshans03N}, and its security against Gaussian individual attacks based on 
entanglement was also demonstrated \cite{Grosshans02R}. Later, it was established that the CV-QKD 
protocol with Gaussian-modulated coherent states is in fact secure against collective attacks \cite{Grosshans05,Christandl04}. 
However the performance of such protocol is severely hindered by a lengthy error correction procedure, specially if the 
signal-to-noise ratio is small \cite{Lodewyck07}. This in practice limits the protocol range to about 30km, although the transmission 
distance can be increased up to $\sim$ 50km, as shown in \cite{Jouguet13}, using a multidimensional reconciliation code, that is, basically 
creating a virtual binary modulation channel. An alternative approach to increase the range of the CV-QKD protocols is to employ
a non-Gaussian (discrete) modulation of coherent states, rather than a Gaussian modulation \cite{Leverrier09,Leverrier11E,Leverrier11}. 
In this case, the encoding is done using a small number of states (e.g. two or four), having fixed amplitudes, which allows for 
a much efficient reconciliation procedure and, consequently, a better performance. Furthermore, it is possible to prove the security of 
a protocol with discrete modulation against general attacks using the already established security proof of the protocol employing coherent 
states with Gaussian modulation \cite{Leverrier09,Leverrier19,Lutkenhaus19}. On the other hand, due to the intrinsic difficulty to implement 
quantum repeaters using light states with Gaussian statistics \cite{Grangier11}, it would be convenient to move away from the Gaussian realm.
A possible way of circumventing such shortcomings and increase the transmission range of a practical CV-QKD system would be using non-Gaussian 
states of light as signal states. Previous studies have shown that there may be advantages if continuous variables non-Gaussian states such as 
phase-coherent states \cite{Becir12} or photon-added-then-subtracted coherent states (PASCS) \cite{Borelli16,Srikara20} are used in place of 
coherent states. Indeed, the robustness of CV-QKD protocols employing these types of non-Gaussian states has been demonstrated for specific 
attacks of the eavesdropper \cite{Becir12,Borelli16,Srikara20}.

In this work, we demonstrate the security of a CV-QKD protocol with a discrete modulation of PASCS against collective attacks.
We will discuss a specific case, the four-state protocol, given that it has a better performance than the two-state protocol, for 
instance. The analysis to be presented here relies on the existence of the unconditional security proof for the ``all-Gaussian protocol", 
i.e., the CV-QKD with Gaussian modulation of Gaussian states \cite{Grosshans05,Christandl04}. We show that the protocol with 
non-Gaussian states (PASCS) not only outperforms the equivalent protocol using coherent states, but is also significantly more robust 
against excess noise in the transmission line.

This paper is organized as follows. In Sec. 2, we introduce the PASCS states and the functioning of the protocol. In Sec. 3, we present the 
security proof against collective attacks. In Sec. 4, we analyze the results regarding the performance of the protocol using PASCS and compare 
them with the protocol using coherent states. In Sec. 5, we discuss and summarize our results. 

\section{Protocol with non-Gaussian (discrete) modulation}

The protocol to be analyzed here is the four-state CV-QKD protocol with discrete modulation \cite{Leverrier09,Hirano03}. 
It works as follows: firstly, Alice chooses one state from a set of four states, say: states $\{|\psi_{+}\rangle$, $|\psi_{+i}\rangle\}$ 
representing bit $1$, and states $\{|\psi_{-}\rangle$, $|\psi_{-i}\rangle\}$ representing bit $0$. In a second step Alice sends 
a light signal prepared in the chosen state to Bob, who randomly measures either the quadrature $X$ or the 
quadrature $Y$ via homodyne detection on the received signal. We consider the reverse reconciliation procedure, 
in which Bob sends side information to Alice in order to complete the process of secret key generation. For instance, if Bob 
obtains the value $X_i$ in his quadrature measurement, he will reveal the absolute value $\vert X_i \vert$ to Alice via a public 
classical channel. At this stage, Alice and Bob share a string of correlated bits. They still need to exchange some more information, 
via the classical channel, to perform error correction and privacy amplification, so that they share a secret key at the end of the 
process.

\subsection{Four-state protocol with photon-added-then-subtracted coherent states (PASCS)}

We are interested in using as signal states the continuous variable, photon-added-then-subtracted coherent states \cite{Pariggi07,Wang11} 
having just one photon added and one photon subtracted. Thus, from an initial coherent state $|\eta\rangle$, we first add one photon to it, 
i.e., $|\phi_A\rangle \propto \hat{a}^\dagger|\eta\rangle$ and then subtract one photon from the resulting state, obtaining the PASCS: 
$|1,1,\eta\rangle \propto \hat{a}|\phi_A\rangle$. This state can be written in the Fock basis as \cite{Wang11}
\begin{equation}
|{1,1,\eta}\rangle=\sum _{k=0}^{\infty} \frac{e^{-|\eta|^2/2} \eta^k (k+1)!}{\sqrt{1+3 |\eta|^2+ |\eta|^4}(k!)^{3/2}} | k \rangle.
\end{equation}

An interesting feature of the state $|1,1,\eta\rangle$ is that it can be expressed as a superposition of a coherent state and a 
photon added coherent state (PACS), or $|1,1,\eta\rangle \propto \hat{a} \hat{a}^\dagger |\eta\rangle \propto 
(1 + \hat{a}^\dagger \hat{a})|\eta\rangle \propto |\eta\rangle + \eta |\phi_A\rangle$. 
In other words, this specific PASCS basically consists as a superposition of a Gaussian state (coherent state) with a 
non-Gaussian component (PACS) weighted by the amplitude $\eta$ \cite{Borelli16}. We emphasize that this feature will be
important in our security analysis to be presented further, as the smaller the $\eta$, the closer the PASCS 
will be of a (Gaussian) coherent state. Besides, the protocol is optimized, i.e., we obtain the maximum possible key 
generation rates precisely for $\eta$ small.

In the prepare and measure version of the protocol, Alice randomly chooses one of the four states 
$\{|\psi_{\pm}\rangle = |1,1,\pm\alpha\rangle, |\psi_{\pm i}\rangle = |{1,1,\pm i\alpha}\rangle\}$ 
($\alpha \in \Re$) and sends it to Bob with probability 1/4. Therefore, the state received by Bob can be represented by 
the following density operator

\begin{eqnarray}
{\rho_4}&=&\frac{1}{4}\big(|1,1,\alpha \rangle \langle 1,1,\alpha| + |1,1,-\alpha \rangle \langle 1,1,-\alpha| \nonumber \\
&&+\,|1,1,i\alpha \rangle \langle 1,1,i\alpha| + |1,1,-i\alpha \rangle \langle 1,1,-i\alpha|\big),\label{rhofourstates}
\end{eqnarray}
that is, a statistical mixture of four PASCS.

\section{Security proof against collective attacks}

The security proof of our protocol with discrete modulation of PASCS follows the same steps as in the protocol using coherent
states \cite{Leverrier09}. This is made possible due to the existence of the well-established unconditional security proof of the 
protocol with Gaussian modulation of Gaussian states, allowing us to calculate an upper bound for the information accessible to Eve while
she executes collective attacks \cite{Grosshans05,Christandl04}. 

In order to perform the security analysis we need to construct the entanglement-based version of the protocol \cite{Leverrier09,Leverrier11E,Grosshans03}. 
This is done by using a purification $\vert \Phi_4 \rangle$ of the state $\rho_4$, that is,
\begin{equation}
|{\Phi_4}\rangle= 
\sqrt{\lambda_0} |\phi_0^* \rangle |\phi_0 \rangle+
\sqrt{\lambda_1} |\phi_1^* \rangle |\phi_1 \rangle+
\sqrt{\lambda_2} |\phi_2^* \rangle |\phi_2 \rangle+
\sqrt{\lambda_3} |\phi_3^* \rangle |\phi_3 \rangle.
\end{equation}

We can then diagonalize the density operator in Equation (\ref{rhofourstates})

\begin{equation}
{\rho_4}= \lambda_0 |\phi_0 \rangle \langle \phi_0| + \lambda_1 |\phi_1 \rangle \langle \phi_1| + \lambda_2 |\phi_2 \rangle \langle \phi_2| + \lambda_3 |\phi_3 \rangle \langle \phi_3|,
\end{equation}

\noindent with corresponding eigenvalues 

\begin{equation}
\lambda_{0}=\frac{e^{-\alpha^2}(3 \alpha^2(- \textrm{sen}{(\alpha^2)}+\textrm{senh}{(\alpha^2)}) - (-1+\alpha^4) \textrm{cos}{(\alpha^2)}+(1+\alpha^4)\textrm{cosh}{(\alpha^2)})} {2 (1+3\alpha^2+\alpha^4)} 
\end{equation}

\begin{equation}
\lambda_{1}=\frac{e^{-\alpha^2}(3\alpha^2 (+ \textrm{cos}{(\alpha^2)}+\textrm{cosh}{(\alpha^2)))} - (-1+\alpha^4) \textrm{sen}{(\alpha^2)}+(1+\alpha^4) \textrm{senh}{(\alpha^2)})}
{2 (1+3\alpha^2+\alpha^4)}
\end{equation}

\begin{equation}
\lambda_{2}=\frac{e^{-\alpha^2}(3 \alpha^2(+ \textrm{sen}{(\alpha^2)}+\textrm{senh}{(\alpha^2)}) + (-1+\alpha^4) \textrm{cos}{(\alpha^2)}+(1+\alpha^4)\textrm{cosh}{(\alpha^2)})} {2 (1+3\alpha^2+\alpha^4)} 
\end{equation}

\begin{equation}
\lambda_{3}=\frac{e^{-\alpha^2}(3\alpha^2 ( - \textrm{cos}{(\alpha^2)}+\textrm{cosh}{(\alpha^2)))} + (-1+\alpha^4) \textrm{sen}{(\alpha^2)}+(1+\alpha^4) \textrm{senh}{(\alpha^2)})}{2 (1+3\alpha^2+\alpha^4)},
\end{equation}

\noindent and eigenvectors

\begin{equation}
|{\phi_k}\rangle= \frac{e^{-\alpha^2/2}}{\sqrt{\lambda_k (1+3 \alpha^2+\alpha^4)}}  \sum_{n=0}^{\infty} \frac{\alpha^{4n+k} (1+k+4n)!}{((4n+k)!)^{3/2}} |4n+k \rangle,
\end{equation}

\noindent
where $k \in \{0,1,2,3\}$.

This allows us to obtain the covariance matrix $\Gamma_{AB}$, between Alice and Bob, represented by

\begin{eqnarray}
\Gamma_{AB}=\left(
\begin{array}{cc}
\langle X_A^2 \rangle \mathbb{I}_2 &  \langle X_A X_B \rangle \sigma_Z \\
\langle X_A X_B \rangle\sigma_Z & \langle X_B^2 \rangle \mathbb{I}_2 \\
\end{array}
\right).
\end{eqnarray}

Because of the symmetry of the state $|\Phi_4\rangle$, we can obtain the following matrix elements 
as well as Alice's modulation variance $V_A$ (in the prepare and measure scheme):

\begin{equation}
\langle X_A^2 \rangle = \langle X_B^2 \rangle = V= 1+V_A,
\end{equation}
with
\begin{equation}
V_A(\alpha)=\frac{2 \alpha ^2 \left(\alpha ^4+5 \alpha ^2+4\right)}   {1+3\alpha^2+\alpha^4}.
\label{variancealice}
\end{equation}

The correlation between Alice and Bob is given by:

\begin{equation}
Z_4= \langle X_A X_B \rangle = \langle \Phi_4| (ab+a^\dagger b^\dagger)| \Phi_4 \rangle,
\end{equation}

\begin{equation}
Z_4(\alpha)=\frac{e^{-2 \alpha ^2} \alpha ^2}{2\left(1+3\alpha ^2+\alpha ^4\right)^2}
\left( \frac{A^2}{\sqrt{\lambda_0\lambda_1}}+
\frac{B^2}{\sqrt{\lambda_1\lambda_2}}+
\frac{C^2}{\sqrt{\lambda_2 \lambda_3}}+
\frac{D^2}{\sqrt{\lambda_3 \lambda_0}} \right),
\end{equation}
where
\begin{eqnarray}
A & = & -\left(\alpha ^4-2\right) \cos \left(\alpha ^2\right)+\left(\alpha ^4+2\right) \cosh \left(\alpha ^2\right)+4 \alpha ^2 \left(\sinh \left(\alpha ^2\right)
-\sin \left(\alpha ^2\right)\right) \nonumber \\
B & = & 4 \alpha ^2 \cos \left(\alpha ^2\right)+4 \alpha ^2 \cosh \left(\alpha ^2\right)-\left(\alpha ^4-2\right) \sin \left(\alpha ^2\right)+\left(\alpha ^4+2\right) \sinh \left(\alpha ^2\right) \nonumber \\
C & = & \left(\alpha ^4-2\right) \cos \left(\alpha ^2\right)+\left(\alpha ^4+2\right) \cosh \left(\alpha ^2\right)+4 \alpha ^2 \left(\sin \left(\alpha ^2\right)+\sinh \left(\alpha ^2\right)\right) \nonumber \\
D & = & -4 \alpha ^2 \cos \left(\alpha ^2\right)+4 \alpha ^2 \cosh \left(\alpha ^2\right)+\left(\alpha ^4-2\right) \sin \left(\alpha ^2\right)+\left(\alpha ^4+2\right) \sinh \left(\alpha ^2\right). \nonumber \\
\end{eqnarray}

\begin{figure}[H]
\includegraphics[width=0.75\textwidth]{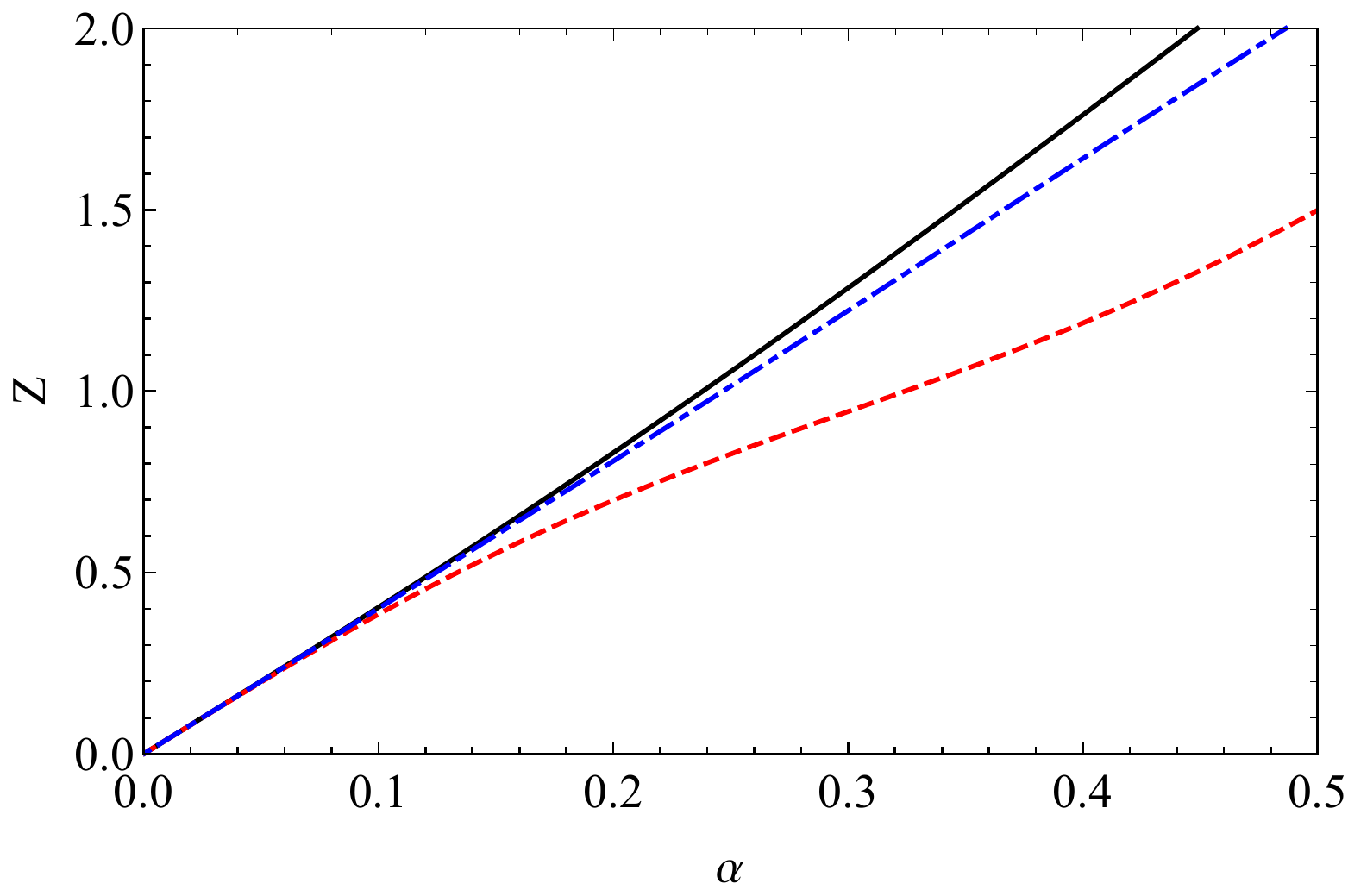}
\caption{Correlations between Alice and Bob modes for the PASCS-based protocol as a function of the amplitude $\alpha$. The solid line 
curve represents the Gaussian modulated protocol ($Z_{Gauss}$), the dashed-doted line curve the four-state protocol ($Z_{4}$), and the 
dashed line curve the two-state protocol ($Z_{2}$). For $\alpha \approx 0.2$ the relative difference between $Z_4$ and $Z_{Gauss}$ is
$\approx 3\%$.}
\label{fig:1}       % Give a unique label
\end{figure}

To ensure the safety of the protocol we have now to find the range of amplitudes $\alpha$ (modulation variances $V_A$) 
in such a way that the resulting covariance matrix is close enough to the corresponding covariance matrix of 
the protocol with Gaussian modulation of PASCS, $\Gamma_{Gauss}$. For instance, we can make a direct comparison between the correlations
$Z_4$ and $Z_{Gauss}$, where $Z_{Gauss}=\sqrt{(1+V_A)^2-1}$. In Fig. \ref{fig:1} we have plotted the function $Z$ as a function of the
amplitude $\alpha$ of the PASCS for the discrete-modulated four-state protocol compared to the Gaussian-modulated protocol. We
have also included a plot of the two-state protocol to show that the four-state protocol is in fact advantageous. We note that 
for small values of the amplitude $\alpha$ the curves $Z_4(\alpha)$ and $Z_{Gauss}(\alpha)$ are nearly indistinguishable.
To be more accurate, we may expand the correlation function up to the third order in $\alpha$,
\begin{equation}
Z_{4}\approx 4 \alpha+4 \left(3 \sqrt{2}-4\right)\alpha^3,
\label{eq:Z4approx}
\end{equation}

\begin{equation}
Z_{Gauss}\approx 4 \alpha+\frac{9}{2}\alpha^3.
\label{eq:Zgaussapprox}
\end{equation}
We have that, for $\alpha \approx 0.2$ the relative difference between $Z_4$ and $Z_{Gauss}$ is just about $\approx 3\%$. 
Besides, the optimum value of the amplitude, $\alpha_{opt} = 0.13$ (the one that maximizes the key rate), lies within the appropriate 
range i.e., $Z_4(\alpha=0.13) \approx Z_{Gauss}(\alpha=0.13)$.

We may carry on the security analysis by assuming that the transmission of the light signals is done via a channel characterized by 
transmissivity $T$ and excess noise $\xi$. The resulting covariance matrix $\gamma_{AB}$ after the transmission is

\begin{eqnarray}
\gamma_{AB}=\left(
\begin{array}{cc}
\gamma_A &  \sigma_{AB} \\
\sigma_{AB} & \gamma_B \\
\end{array}
 \right),
\end{eqnarray}
where 
\begin{eqnarray}
\gamma_{A} & = & (1 + V_{A}) \mathbb{I}_2 \nonumber\\
\gamma_{B} & = & (T V_A+1-T \xi)\mathbb{I}_2  \nonumber\\
\sigma_{AB} & = & Z\sqrt{T} \sigma_{z}.
\end{eqnarray} 

Following the transmission through the non-ideal quantum channel, Bob performs a homodyne detection on the received signal, 
represented by the transformations:

\begin{equation}
\gamma_{A|B}^{hom}=\gamma_{A}-\sigma_{AB} {(X \gamma_{B} X)}^{MP} {\sigma_{AB}}^T,
\end{equation}
where $MP$ is the Moore-Penrose pseudo-inverse and $X = diag(1,0) $.

The resulting (reduced) covariance matrix after Bob's measurement is
\begin{eqnarray}
\gamma_{A|B}^{hom}=\left(
\begin{array}{cc}
V_A+1-\frac{T Z^2}{T V_A+1-T \xi} &  0 \\
0 & V_A+1\\
\end{array}
\right).
\end{eqnarray}
We may now calculate $I_{AB}$, the Shannon mutual information \cite{Shannon48} between Alice and Bob's data,
\begin{equation}
I_{AB}=\frac{1}{2} \log \left(\frac{V_A}{V_{A|B}} \right),
\label{eq:iab}
\end{equation}
\noindent
being $V_{A}$ the modulation variance in Eq. (\ref{variancealice}), and $V_{A|B}$ the conditional quadrature variance 
\cite{Zhao20,Wang19}, which is equal to the first diagonal element of the conditional matrix $\gamma_{A|B}^{hom}$, 

\begin{equation}
V_{A|B}=V_A+1-\frac{T Z^2}{T V_A+1-T \xi}.
\label{eq:vabarrab}
\end{equation}

The the upper bound on the information Eve can obtain by carrying out a collective attack is given by $S_{BE}$,
the Holevo information \cite{Grosshans05,Christandl04}. For the PASCS, this quantity is written as

\begin{equation}
S_{BE}=G\left(\frac{\nu_1-1}{2}\right)+G\left(\frac{\nu_2-1}{2}\right)-G\left(\frac{\nu_3-1}{2}\right),
\label{eq:holevoinf}
\end{equation}
with

\begin{equation}
G(x)=(x+1) \log(x+1) - (x) \log(x),
\end{equation}

\begin{equation}
\nu_{1}=\sqrt{\frac{1}{2}\left(\Delta+\sqrt{\Delta^2-4\delta} \right)},
\end{equation}

\begin{equation}
\nu_{2}=\sqrt{\frac{1}{2}\left(\Delta-\sqrt{\Delta^2-4\delta} \right)},
\end{equation}

\begin{equation}
\nu_3=\sqrt{(V_{A}+1)\left((V_{A}+1)-\frac{T Z_{4}^2}{\xi T+TV_{A}+1}\right)},
\end{equation}

\begin{equation}
\Delta=\xi ^2 T^2+\left(T^2+1\right) V_{A}^2+2 V_{A} \left(\xi T^2+T+1\right)+2 \xi T-2 T Z_{4}^2+2,
\end{equation}
and

\begin{equation}
\delta=\left(T V_{A}^2+V_{A} (\xi  T+T+1)+T \left(\xi -Z_{4}^2\right)+1\right)^2.
\end{equation}

If we use now the expressions in Eqs. (\ref{variancealice}), (\ref{eq:vabarrab}), (\ref{eq:iab}) and (\ref{eq:holevoinf}), we
obtain the secret key rate $K$ under collective attacks, given by: 
\begin{equation}
K= \beta I_{AB} - S_{BE},
\end{equation}
where $\beta$ is the reconciliation efficiency of the protocol. The key rate $K$ is the quantity we will use below to 
analyze the performance of our PACS-based protocol.

\section{Results}

\noindent
Now we would like to present our results concerning the performance of a CV-QKD protocol using PASCS and also compare them to the 
protocol with coherent states. We recall that we have used the optimum value of the amplitude, $\alpha_{opt}=0.13$, that maximizes 
the key rate. This corresponds to a relative difference between $Z_4$ and $Z_{Gauss}$ of $\approx 1.3\%$ (see Fig. \ref{fig:1} ) 
and Eqs. (\ref{eq:Z4approx}) and (\ref{eq:Zgaussapprox}).

\begin{figure}[H]
\includegraphics[width=0.75\textwidth]{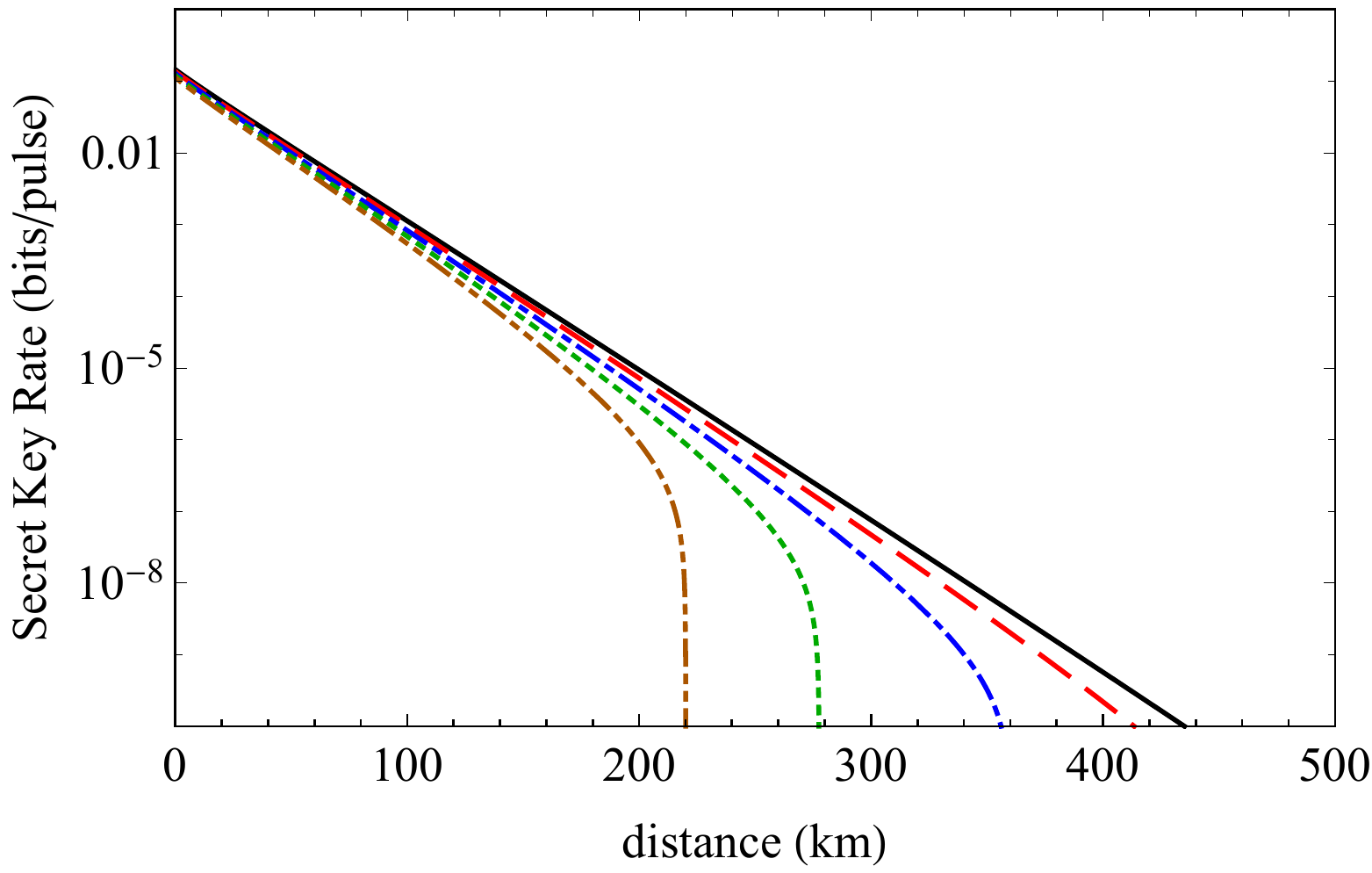}
\caption{Secret key rate of the PASCS-based four-state protocol with homodyne detection, photodetector quantum efficiency of $100\%$ and 
perfect reconciliation efficiency ($\beta = 100\%$). From right to left, the excess noise $\xi$ is 0.002, 0.004, 0.006, 0.008 
and 0.01, and the optimum amplitude is $\alpha_{opt} =  0.13$ in all plots.}
\label{fig:2}       % Give a unique label
\end{figure}

In Fig. \ref{fig:2} it is shown the key generation rate for the ideal case, that is, with Bob performing homodyne detection with $100\%$ of efficiency 
and a perfect reconciliation rate $(\beta = 100\%)$. The graph shows the influence of excess noise on the key rate and transmission distance. 
We note that for very low values of excess noise, $\xi = 0.002$ and $\xi = 0.004$, we can have a secure key generation 
(at a rate $K \approx 10^{-10}$ bits/pulse) with a transmission range exceeding $400$ km before saturation.

\begin{figure}
\includegraphics[width=0.75\textwidth]{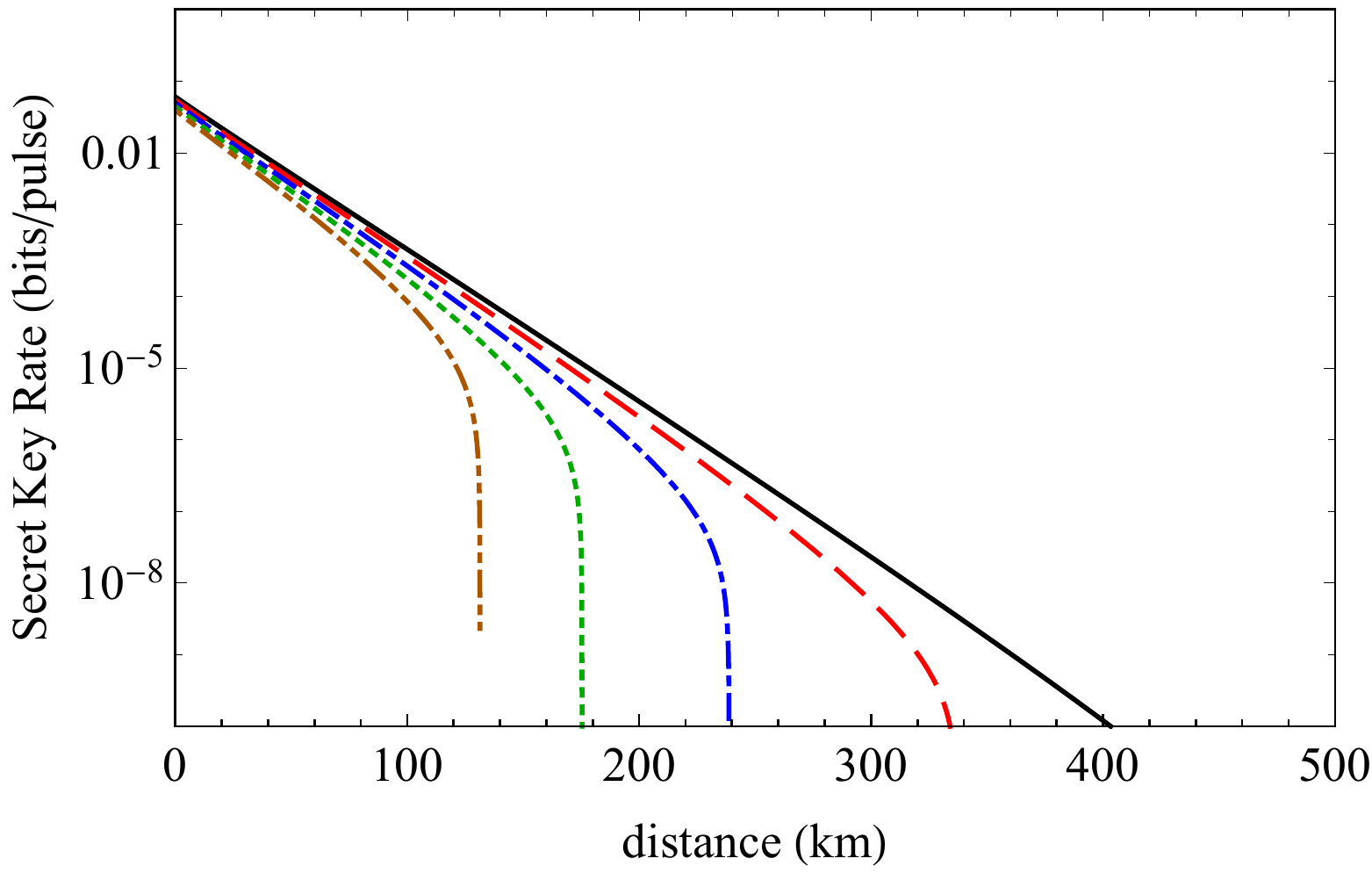}
\caption{Secret key rate of the PASCS-based four-state protocol with homodyne detection, photodetector quantum efficiency of $60\%$ and 
imperfect reconciliation efficiency ($\beta = 80\%$). From right to left, the excess noise $\xi$ is 0.002, 0.004, 0.006, 0.008 
and 0.01, and the optimum amplitude is $\alpha_{opt} =  0.13$ in all plots.}
\label{fig:3}       % Give a unique label
\end{figure}

In Fig. \ref{fig:3} we analyze a more realistic case where we consider efficiencies of $\beta = 80\%$ for the reconciliation, and of $60\%$ for
the photodetector. There is a drop in the transmission distance for a given key rate (as expected), and this gets worse for larger values of
$\xi$ (excess noise), as can be clearly seen in Fig. \ref{fig:3}.

\begin{figure}
\includegraphics[width=0.75\textwidth]{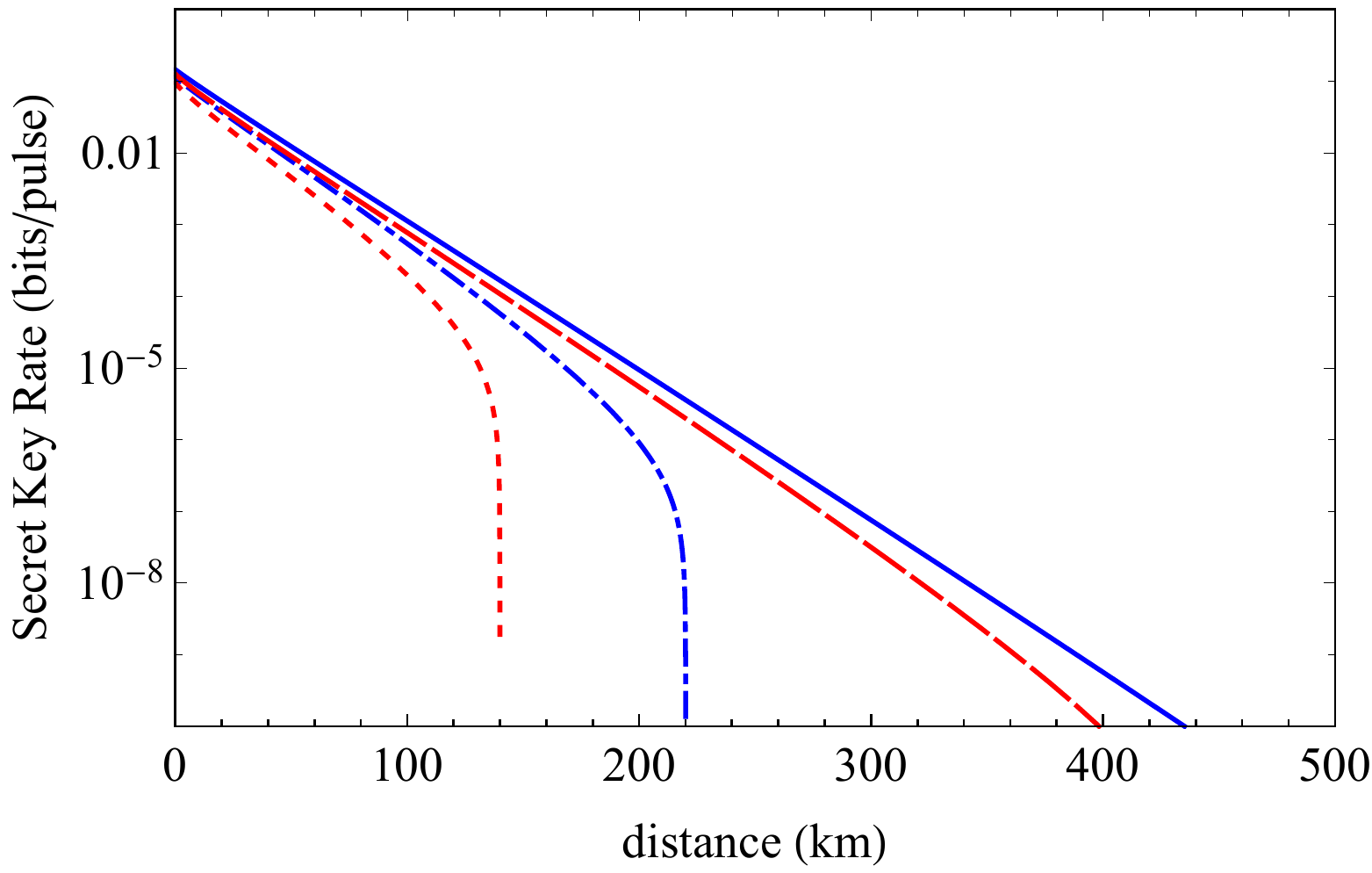}
\caption{Comparison between the key generation rates from the protocol using PASCS with that of the protocol using coherent states.
Both are four-state protocols with homodyne detection, photodetector quantum efficiency of $100\%$ and 
perfect reconciliation efficiency ($\beta = 100\%$). From right to left, excess noise is $0.002$ (solid line for the PASCS and dashed line 
for the coherent states) and $0.01$ (dashed-doted line PASCS and dotted line for the coherent states). The optimum amplitudes 
used are $\alpha_{opt}^{(\mbox{\tiny{PASCS}})} = 0.13$ and $\alpha_{opt}^{(coh)} = 0.25$, for the PASCS and coherent states, respectively.}
\label{fig:4}       % Give a unique label
\end{figure}

We may also compare the performances of the protocol using PASCS with the one using coherent states \cite{Leverrier09}, having
optimum modulation amplitudes of $\alpha_{opt}^{(\mbox{\tiny{PASCS}})} = 0.13$ and $\alpha_{opt}^{(coh)} = 0.25$, respectively. 
In Fig. \ref{fig:4} we have plotted the key rate as a function of the transmission distance for perfect reconciliation $(\beta = 100\%)$ 
and photodetector efficiency of $100\%$. This is done for both low excess noise $(\xi = 0.002)$ and higher excess noise $(\xi = 0.01)$. 
Remarkably, the PASCS-based protocol has a significantly superior performance if the excess noise is higher, as shown in the two curves 
on the left in Fig. \ref{fig:4}, although they do not differ much in the low noise case (the two curves on the right). For the
protocol with coherent states the curve saturates at a transmission distance of $\approx 140$ km ($K \approx 10^{-5}$ bits/pulse), 
while for the PASCS this occurs at a transmission distance of $\approx 220$ km ($K \approx 10^{-7}$ bits/pulse).

\begin{figure}
\includegraphics[width=0.75\textwidth]{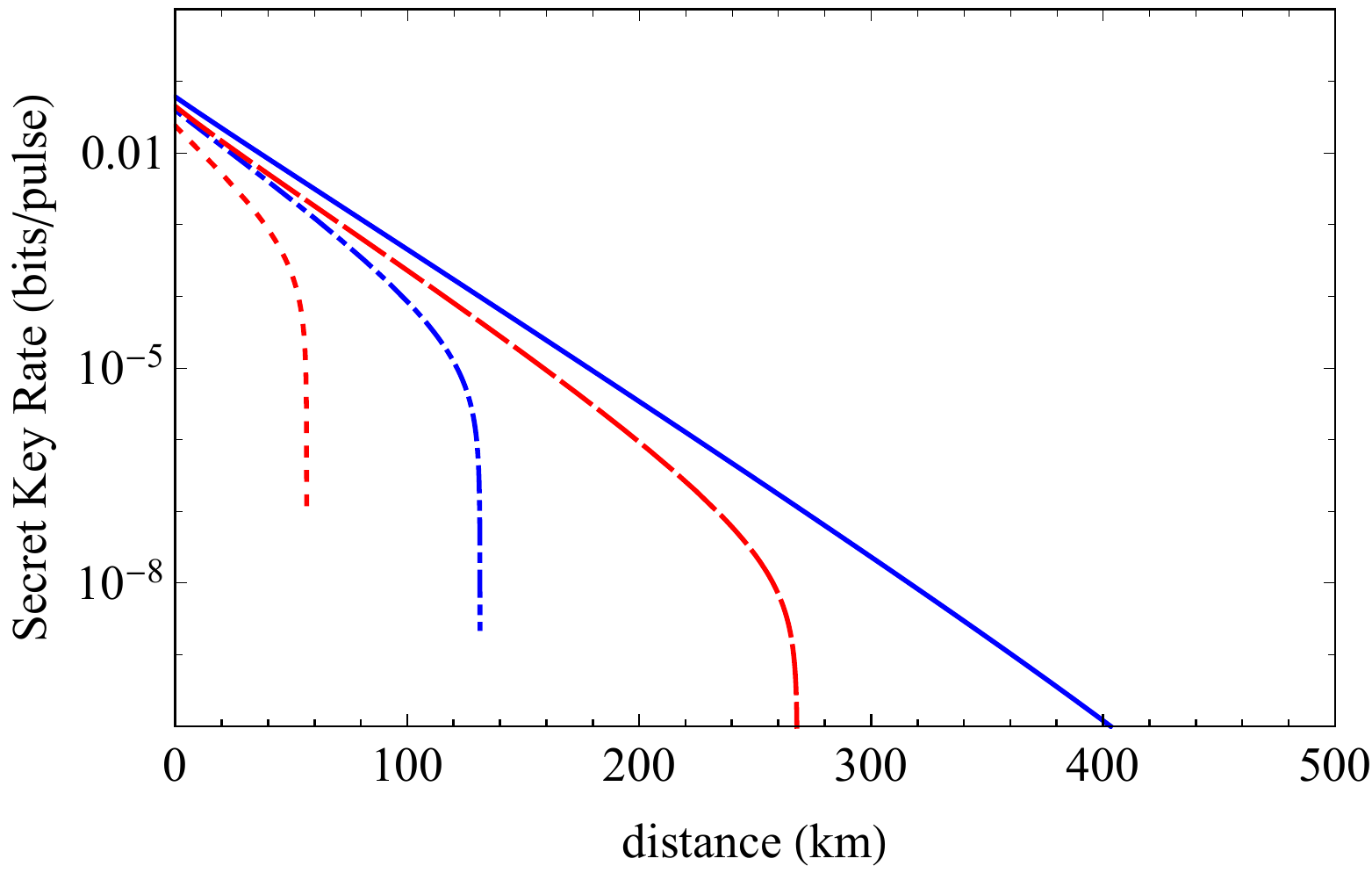}
\caption{Comparison between the key generation rates from the protocol using PASCS with that of the protocol using coherent states.
Both are four-state protocols with homodyne detection, photodetector quantum efficiency of $60\%$ and 
imperfect reconciliation efficiency ($\beta = 80\%$). From right to left, excess noise is $0.002$ (solid line for the PASCS and dashed line 
for the coherent states) and $0.01$ (dashed-doted line PASCS and dotted line for the coherent states). The optimum amplitudes 
are the same as in Fig. \ref{fig:4}.}
\label{fig:5}       % Give a unique label
\end{figure}

If the conditions in Bob's station are non-ideal, e.g., with a reconciliation coefficient $\beta = 80\%$ and photodetector 
efficiency of $60\%$, the performance of the protocols will be of course degraded. However, the use of PASCS in place of coherent 
states remains advantageous, as the former still allows for higher key generation rates and a longer transmission range
especially for high excess noise, as shown in the two curves on the left in Fig.\ref{fig:5}.

\newpage
\section{Conclusion}
We presented an analysis of the performance of a CV-QKD protocol using as signal states photon-added-then-subtracted 
coherent states (PASCS). We computed the secret key generation rate against collective attacks for a protocol with a discrete 
modulation of four states. The PASCS can be written as a quantum superposition of a coherent state $\vert \alpha\rangle$ and a 
photon-added coherent state multiplied by the amplitude $\alpha$. Thus, despite being non-Gaussian states, they become ``close" to
the (Gaussian) coherent states for small amplitudes $\alpha$, which justifies the calculation of the upper bound of the
Holevo information between Eve and Bob based on the result obtained for the protocol with Gaussian modulation. 
We compared our results (secret key rates) with the ones obtained for the four-state protocol using coherent states and 
in every scenario we studied, the PASCS-based protocol outperforms the coherent state-based one. In particular, we found that
the difference between the maximum transmission distances achieved in each protocol (PASCS vs. coherent states) becomes larger 
for noisier transmission lines (large $\xi$). In other words, the protocol with photon-added-then-subtracted coherent states is
considerably more robust against excess noise than the corresponding one using coherent states. Our work is a step towards the 
development of CV-QKD protocols using non-Gaussian states, aiming at the improvement of the efficiency of quantum cryptography 
systems.  

\section*{Acknowledgments}
This work has been supported by Conselho Nacional de Desenvolvimento Científico e Tecnol\'ogico, 
(CNPq) Brazil, via the Instituto Nacional de Ci\^encia e Tecnologia - Informa\c c\~ao qu\^antica
 (INCT-IQ), grant N${\textsuperscript{\underline{o}}}$ 465469/2014-0.

\end{document}